# Time-dependent quantum Monte Carlo and the stochastic quantization

Ivan P. Christov

Physics Department, Sofia University, 1164 Sofia, Bulgaria

**Abstract**

We examine the relation between the recently proposed time-dependent quantum Monte Carlo (TDQMC) method and the principles of stochastic quantization. In both TDQMC and stochastic quantization particle motion obeys stochastic guidance equations to preserve quantum equilibrium. In this way the probability density of the Monte Carlo particles corresponds to the modulus square of the many-body wave function at all times. However, in TDQMC the motion of particles and guide waves occurs in physical space unlike in stochastic quantization where it occurs in configuration space. Hence the practical calculation of time evolution of many-body fully correlated quantum systems becomes feasible within the TDQMC methodology. We illustrate the TDQMC technique by calculating the symmetric and antisymmetric ground state of a model one-dimensional Helium atom, and the time evolution of the dipole moment when the atom is irradiated by a strong ultrashort laser pulse.

# 1. Introduction

Recently, time-dependent quantum dynamics of complex systems have attracted much attention connected with the advent of sources of femto- and atto-second duration laser pulses and the opportunities these offer for time-domain studies of molecules, clusters, and solids. There has been an increasing need to develop new numerical tools capable of carrying out reliable time-dependent calculations for the interaction of such systems with external fields. The conventional approaches for solving the time dependent Schrödinger equation (TDSE), such as basis set expansion techniques, require computational effort that scales exponentially with the number of particles involved, that prohibits their implementation for realistic conditions. On the other hand, the known semiclassical methods allow to calculate only a part of the effects that accompany the interaction of atoms and small to medium size quantum systems with ultrashort laser pulses. In order to comprehend the many-body quantum problem, approximations are introduced. One of the most popular of those is the density functional theory (DFT),[1] which is a typical mean-field theory where the detailed fluctuating forces between the electrons are replaced by an averaged force. In case of $N$ electrons, this successfully reduces the dimensionality of the Schrödinger equation from $3N$ to three, for each electron. Although exact in principle, DFT involves an unknown exchange-correlation functional which is usually approximated. For time-dependent studies, DFT is upgraded to time-dependent DFT (TDDFT),[2] where the exchange-correlation potential is estimated using the adiabatic local-density approximation (A-LDA)[3]. Although A-LDA has proven to work well for excitations in atoms and molecules, it fails in other important cases that involve significant deformations of the electron density such as double ionization and charge transfer [4,5]. It has been shown that TDDFT suffers from "memory effects" in that the time-dependent density functionals remember both the entire history of the density and the initial wave function [6,7]. Other failures of TDDFT have also been reported, which are related to states with large angular momentum [8]. Other time-dependent mean-field methods combine DFT with quantum fluid dynamics that use parameterized exchange-correlation potentials within LDA [9]. More reliable, but computationally more expensive is the multiconfiguration time dependent Hartree-Fock method [10]. Another method to treat correlated many-body problems employs time dependent configuration-interaction [11].

An alternative approach to quantum many-body systems, which is principally different from the above methods, is offered by the quantum Monte Carlo (QMC) techniques [12,13]. The quantum Monte Carlo methods are stochastic methods in obtaining the expectation values of quantum stationary states and their energies while keeping favorable scaling with the system dimensionality. That scaling is typically linear or low-order-polynomial (no worse than $N^3$) which, combined with the intrinsic parallelism of the QMC algorithms and the fast increasing computer power, allows their robust application to large molecules, nano-structures, and condensed-matter systems. Diffusion Quantum Monte Carlo (DMC) employs random particles (walkers) whose probability distribution in configuration space is close to the modulus of the many-body wave function, assuming that the wave-function is positive everywhere. The evolution of the walker distribution towards the ground state of the system is based on the similarity between the imaginary time Schrödinger equation and a generalized diffusion process, where the kinetic energy term of the Schrödinger equation corresponds to random diffusive jumps of the walkers while the potential term leads to birth/death of walkers [14]. Therefore, the process of walkers' evolution towards the ground state of the system is represented as a combination of diffusion and branching processes, where the number of diffusing walkers increases or decreases at a given point proportional to the potential energy at that point. One limitation of the random walk QMC is that it can be very inefficient due to the uncontrolled branching process. If the potential becomes large and negative, as it is for Coulomb potential, the number of copies of a walker increases unrestrictedly, which leads to a large fluctuations in the estimate of the ground state energy. A remedy of that problem is offered by introducing an importance sampling technique. The idea is to make an initial guess for the many-body wave function, named as guide function $\Psi_G(\mathbf{R})$, to guide the walkers toward the most important regions of the potential. In fact, the importance sampling causes bias in the diffusion process, which directs the walkers towards parts of configuration space where $|\Psi_G(\mathbf{R})|$ is large. By using a sufficiently accurate guide function, the number of branching events can be significantly reduced together with the statistical error in the final result. Such guide functions are usually produced by Hartree-Fock or variational Monte Carlo method. A common drawback of the present day QMC methods is that they cannot describe time-dependent processes.

Recently, a new time dependent quantum Monte Carlo (TDQMC) technique was introduced which offers important advantages compared to the conventional QMC[15,16]. TDQMC uses walkers guided by de Broglie-Bohm pilot waves, where a separate guide wave is associated with each individual walker. The most significant advantage of TDQMC is that it allows one to carry out time dependent studies of quantum objects and their interaction with external electromagnetic fields, while preserving fully correlated quantum dynamics. At the same time, unlike in DMC, in TDQMC the distribution of the walkers in configuration space corresponds to the quantum probability density, and therefore this method is not sensitive to the sign of the many-body wavefunction (known as fermion sign problem in conventional DMC). Since in TDQMC the guide functions evolve together with the particle configurations, no initial guess for both the many-body wave-function and for the ground state energy is needed. Moreover, in TDQMC all calculations are performed in physical space for both the particles (walkers) and the associated guiding waves, while the particle distribution in configuration space and the corresponding probability density function are calculated only when needed. Since the guide functions evolve concurrently with the walker's positions, the branching is further reduced in TDQMC as compared to importance-sampling DMC. We note that the de Broglie-Bohm methodology has already been a subject of extensive research for potential applications in quantum chemistry [17-22]. In that work the quantum trajectories are determined usually by solving the quantum hydrodynamic equations for different conditions. However, these equations are nonlinear and contain quantum potentials which can be an obstacle for their robust numerical treatment. Instead, in TDQMC coupled single-particle Schrödinger equations are solved, which relaxes the convergence and stability requirements for states with static and/or time-dependent nodes. Other work uses Monte Carlo sampling of initial conditions for de Broglie-Bohm trajectories to calculate the trajectories of particles for simple Hamiltonians [23].

In this work the connection between the TDQMC technique and the stochastic quantum mechanics that uses de Brogie-Bohm (dBB) trajectories is studied. It is shown that TDQMC incorporates elements of the dBB formulation of stochastic quantum mechanics into the quantum Monte Carlo framework, where the correlated electron motion is accounted for *ab initio* using explicit Coulomb potentials, instead of using exchange-correlation potentials.

## 2. Guide-function DMC

We start with a brief description of the guide-function DMC algorithm [12,13]. The basis of DMC is to write the Schrödinger equation:

$$i\hbar \frac{\partial}{\partial t}\Psi(\mathbf{R},t) = -\frac{\hbar^2}{2m}\nabla^2\Psi(\mathbf{R},t) + V(\mathbf{R})\Psi(\mathbf{R},t) \qquad (1)$$

in imaginary time $\tau = it$, which yields:

$$\hbar \frac{\partial}{\partial \tau}\Psi(\mathbf{R},\tau) = \frac{\hbar^2}{2m}\nabla^2\Psi(\mathbf{R},\tau) + [E_T - V(\mathbf{R})]\Psi(\mathbf{R},\tau), \qquad (2)$$

where $\mathbf{R} = (\mathbf{r}_1, \mathbf{r}_2, ..., \mathbf{r}_N)$ is a 3$N$ dimensional vector which specifies the coordinates of $N$ electrons, $\nabla = (\nabla_1, \nabla_2, ..., \nabla_N)$, and we have introduced an energy offset $E_T$. The imaginary time Schrödinger equation (Eq. 2) resembles a diffusion equation in 3$N$-dimensional configuration space, and we proceed with the short time approximation of the Green function of that equation, where $|\Psi(\mathbf{R},\tau)|$ plays a role of density of the diffusing particles (walkers), and the term $[E_T - V(\mathbf{R})]\Psi(\mathbf{R},\tau)$ describes the "branching" or creation/annihilation of walkers. Thus, the entire process of walker evolution can be described as a combination of diffusion and branching events, in which the number of walkers at a given point is proportional to $|\Psi(\mathbf{R},\tau)|$. In quantum Monte Carlo, the updated random positions of the walkers are generated using Markov process where each move is independent of the previous history of the system. The reason for using Markov process is that after sufficiently long time the most likely state of the system is established, similarly to reaching equilibrium distribution in thermodynamics. The energy $E_T$ can be adjusted so that the fluctuations of the overall number of walkers can be restricted around some prescribed value. The efficiency of the algorithm is significantly improved if importance sampling is used. To this end, an auxiliary guiding function $\Psi_G(\mathbf{R})$ is introduced, where we assume that the particle distribution is given by the product $f(\mathbf{R},\tau) = \Psi_G(\mathbf{R})\Psi(\mathbf{R},\tau)$. Then, the new function $f(\mathbf{R},\tau)$ satisfies the master equation for a system of Brownian particles undergoing a stochastic diffusion process (Fokker-Planck equation):

$$\hbar \frac{\partial}{\partial \tau} f(\mathbf{R},\tau) = \frac{\hbar^2}{2m} \nabla^2 f(\mathbf{R},\tau) - \hbar \nabla \cdot [\mathbf{v}_G(\mathbf{R}) f(\mathbf{R},\tau)] + [E_T - E_L(\mathbf{R})] f(\mathbf{R},\tau) \ , \qquad (3)$$

where:

$$\mathbf{v}_G(\mathbf{R}) = \frac{\hbar}{m} \frac{\nabla \Psi_G(\mathbf{R})}{\Psi_G(\mathbf{R})} \qquad (4)$$

is interpreted in QMC as "quantum force" (with a dimension of velocity) whose role is to enhance the density of walkers in the regions where that density is large and vice versa, and $E_T - E_L(\mathbf{R})$ is the branching term, which is defined in terms of the local energy $E_L(\mathbf{R})$:

$$E_L(\mathbf{R}) = V(\mathbf{R}) - \frac{\hbar^2}{2m} \frac{1}{\Psi_G(\mathbf{R})} \nabla^2 \Psi_G(\mathbf{R}) \qquad (5)$$

Algorithmically, first we initialize a set of configurations with a probability density distribution close to $|\Psi_G(\mathbf{R})|^2$. Then, the *k-th* walker from the *i-th* electron ensemble makes moves as a Brownian particle according the stochastic equation (Wiener type of process):

$$d\mathbf{r}_i^k = \mathbf{v}_G(\mathbf{r}_i^k) dt + \boldsymbol{\eta} \sqrt{\frac{\hbar}{m}} dt , \qquad (6)$$

where $\mathbf{r}_i^k$ is the current coordinate of the walker, and $\boldsymbol{\eta}$ is a vector random variable with zero mean and unit variance (Gaussian white noise). Next, the acceptance probability and the branching probability are estimated and the move of the walker is accepted or copies of that walker are made. In case of guide function with nodes, we check whether the move has caused the walker to cross a nodal surface. The nodal surfaces correspond to the zeroes of the many-body wavefunction, where it changes sign. If this has occurred, we reject the move and go to the next particle. For fermions, the preliminary chosen guiding function $\Psi_G(\mathbf{R})$ determines the position of the nodes of the ground-state wavefunction to ensure that $f(\mathbf{R},\tau)$ is always of the same sign. Therefore, guide function with exact positions of the nodes of the quantum state is required, which is a major impediment in DMC since the nodes of the fermionic wave-function are generally unknown. This problem is remedied in TDQMC, as shown below.

## 3. The de Broglie-Bohm theory within Monte Carlo context

Quantum Monte Carlo is not the only approach to associate quantum mechanics with stochastic process. Another possible formulation of quantum mechanics in terms of ensembles of trajectories is provided by the de Broglie-Bohm theory and its modifications [24-26]. In the original dBB theory, it is assumed that the quantum-mechanical system consists of waves and point particles which are guided by these waves. The particle concept is introduced by representing the wave-function as a polar decomposition:

$$\Psi(\mathbf{R},t) = R(\mathbf{R},t)\exp[iS(\mathbf{R},t)/\hbar] \quad , \qquad (7)$$

where $R(\mathbf{R},t)$ and $S(\mathbf{R},t)$ are real-valued functions of space and time. Then, inserting Eq. (7) into Eq. (1) and separating the real and imaginary parts, we obtain the two equations of quantum hydrodynamics [26]:

$$\frac{\partial S(\mathbf{R},t)}{\partial t} + \frac{1}{2m}\sum_{i=1}^{N}[\nabla_i S(\mathbf{R},t)]^2 + Q(\mathbf{R},t) + V(\mathbf{R},t) = 0 \qquad (8)$$

$$\frac{\partial R^2(\mathbf{R},t)}{\partial t} + \frac{1}{m}\sum_{i=1}^{N}\nabla_i \cdot (R^2(\mathbf{R},t)\nabla_i S(\mathbf{R},t)) = 0, \qquad (9)$$

where:

$$Q(\mathbf{R},t) = -\frac{\hbar^2}{2m}\sum_{i=1}^{N}\frac{\nabla_i^2 R(\mathbf{R},t)}{R(\mathbf{R},t)} \qquad (10)$$

is the many-body quantum potential. Equation (8) represents a generalized Hamilton-Jacobi equation, whereas Eq. (9) is a continuity equation for the probability density. Next, we apply the operator $\nabla_j \equiv \partial/\partial \mathbf{r}_j$ to both sides of Eq. (8), and after rearranging we get:

$$\left\{\frac{\partial}{\partial t} + \frac{1}{m}\left[\sum_{i=1}^{N}\nabla_i S(\mathbf{r}_1,...,\mathbf{r}_N,t)\cdot\nabla_i\right]\right\}\nabla_j S(\mathbf{r}_1,...,\mathbf{r}_N,t) = -\nabla_j[Q(\mathbf{R},t)+V(\mathbf{R},t)] \qquad (11)$$

Now, we set the velocity fields through the de Broglie-Bohm pilot-wave relation:

$$\frac{d\mathbf{r}_i}{dt} = \frac{1}{m}\nabla_i S(\mathbf{r}_1,...,\mathbf{r}_N,t)_{\mathbf{r}_j=\mathbf{r}_j(t)}, \qquad i,j=1,...,N \qquad (12)$$

which transforms Eq. (11) into a second-order Newtonian type equation:

$$m\frac{d^2\mathbf{r}_i}{dt^2} = \{-\nabla_i[Q(\mathbf{r}_1,...,\mathbf{r}_N,t)+V(\mathbf{r}_1,...,\mathbf{r}_N,t)]\}_{\mathbf{r}_j=\mathbf{r}_j(t)}, \qquad i,j=1,...,N \qquad (13)$$

In the standard dBB theory, each particle follows a well defined trajectory $r_i(t)$ which is determined by the initial particle positions $r_i(t=0)$ and by one of the two equations of motion, Eq. (12) or Eq. (13). Equation (12) only involves the first derivative of particle position while Eq. (13) is similar to Newtonian equation with force $F_i = -\nabla_i [Q(r_1,...,r_N,t) + V(r_1,...,r_N,t)]$. Since particle motion can be determined by directly solving Eq. 13, much attention has been focused on the evaluation of the quantum force. However, it is seen from Eqs. (10) and (13) that the quantum force is given by the third derivative of the instantaneous particle density, which makes its accurate numerical estimation difficult. In order to evaluate the derivatives, moving weighted least squares fitting schemes [27,28] or Gaussian expansions [29] have been used.

The stochastic (Monte Carlo) interpretation of dBB theory originates from the lack of preliminary knowledge of the precise initial positions of the particles, so we have to use a statistical ensemble of particles. However, since the particle distribution $P(\mathbf{R},t)$ is not necessarily connected with the probability density given by the wave-function $|\Psi(\mathbf{R},t)|^2$, the equality $P(\mathbf{R},t) = |\Psi(\mathbf{R},t)|^2$ is often considered to be one of the postulates in dBB theory [30-32]. In order to further clarify this point, Bohm introduced a random component in particles' motion where each particle interacts with other (external to the system) particles, e.g. via collisions [33]. As a result of that random motion $P(\mathbf{R},t)$ will finally converge to $|\Psi(\mathbf{R},t)|^2$. Also, Bohm and Vigier [34] have developed statistical mechanics of particles, where independently of the choice of $P(\mathbf{R},t=0)$, $P(\mathbf{R},t)$ will tend asymptotically to $|\Psi(\mathbf{R},t)|^2$ owing to random fluctuations. It has been shown that, under these assumptions, the predictions of the dBB theory must agree with those of the orthodox quantum theory [35]. Similar conclusions, but from a different perspective, have been drown,[36] where is has been shown that the $|\Psi(\mathbf{R},t)|^2$ distribution in Bohmian mechanics corresponds to an 'equilibrium' distribution, similar to the thermodynamical equilibrium in statistical mechanics. Therefore, in the stochastic interpretation of dBB theory, an arbitrary quantum system tends to quantum equilibrium via fluctuations, where at equilibrium we have $P(\mathbf{R},t) = |\Psi(\mathbf{R},t)|^2$. Then, Eq. 12 has to be replaced by a stochastic guidance equation, where each particle has a mean velocity $\langle \mathbf{v}_i \rangle = \nabla_i S / m$, and a stochastic component is added to the particle motion.

Another closely related theory is due to Nelson [37] and others, [38,39] who showed that the stochastic quantization can be represented by the following guidance equation for the dBB trajectories:

$$d\mathbf{r}_i^k = \mathbf{v}(\mathbf{r}_i^k)dt + \mathbf{\eta}\sqrt{\frac{\alpha\hbar}{m}dt}, \qquad (14)$$

where:

$$\mathbf{v} = \frac{1}{m}\nabla S + \alpha \frac{\hbar}{2m}\frac{\nabla|\Psi|^2}{|\Psi|^2} = \frac{\hbar}{m}(\alpha\,\text{Re} + \text{Im})\frac{\nabla\Psi}{\Psi}, \qquad (15)$$

and $\boldsymbol{\eta}$ is a random variable with zero mean and unit variance. Then, the parameter $\alpha = 1$ corresponds to Nelson stochastic mechanics, while $\alpha = 0$ leads to conventional dBB mechanics (Eq. 12). The velocity $\mathbf{v}=\mathbf{v}_D+\mathbf{v}_O$ is a sum of the drift ($\mathbf{v}_D$) and osmotic ($\mathbf{v}_O$) velocities, where the drift velocity is given by:

$$\mathbf{v}_D = \frac{1}{m}\nabla S(\mathbf{R},t) = \frac{\hbar}{m}\mathrm{Im}\frac{\nabla\Psi(\mathbf{R},t)}{\Psi(\mathbf{R},t)} \quad , \tag{16}$$

and the osmotic velocity is:

$$\mathbf{v}_O = \alpha\frac{\hbar}{2m}\frac{\nabla|\Psi(\mathbf{R},t)|^2}{|\Psi(\mathbf{R},t)|^2} = \alpha\frac{\hbar}{m}\mathrm{Re}\frac{\nabla\Psi(\mathbf{R},t)}{\Psi(\mathbf{R},t)} \tag{17}$$

It is seen from Eq.17 that the osmotic velocity pushes the particles to the regions where $|\Psi(\mathbf{R},t)|^2$ is large, and keeps them away from the nodes of $|\Psi(\mathbf{R},t)|^2$, similarly to the action of the importance sampling in DMC. It can be shown that there is an equilibrium state in which the osmotic velocity is balanced by the diffusion current so that the mean velocity of the particle is $\langle\mathbf{v}_i\rangle = \nabla_i S/m$. In fact, Eq. (14) represents a Langevin equation which describes the particle motion to be a result of drift and diffusion processes which can be derived on the basis of more general assumptions from classical stochastic dynamics (not related to dBB theory) [40]. Other approaches include complex velocities,[41,42] and Parisi-Wu stochastic quantization [43].

## 4. Time dependent quantum Monte Carlo (TDQMC)

In TDQMC method some similarities between DMC and the stochastic quantum dynamics with dBB trajectories are employed. It is seen from Eq. (6) and Eq. (14) that the stochastic processes of random walk are practically identical for the two theories. The guidance equations, Eq. (4) and Eq. (15), are also very close. These similarities suggest that Monte Carlo simulations can be used for dBB trajectories, not only for the initial ensemble of particles, but for the whole time evolution of the system. The physical quantities of interest can be calculated as ensemble averages over the evolved distribution, using Monte Carlo integration. Reasons why the standard DMC methodology is not appropriate for time-dependent studies include the following:

1. In DMC we would have to calculate the real-time evolution of the many-body state by solving the full time-dependent Schrödinger equation (Eq. (1)), or alternatively by fitting the probability density distribution in configuration space for each instant of time [22], and then use Eq. (12) or Eq.(13) to calculate the particle motion. Clearly, that would be prohibitively time-expensive even for small systems.

2. In DMC the symmetry of the wavefuncton under exchange of the electrons and the positions of the nodes are pre-assigned by the chosen guiding function, which is usually represented in a Slater-Jastrow form [44]. Instead, in TDQMC we use guide functions which evolve in time, concurrently with the evolution of the dBB particles (walkers), that implies a natural evolution of the nodes.

In order to be able to build a numerically tractable algorithm for calculating quantum dynamics we have to address point 1 above. Since in TDQMC we treat symmetrically the particles and the guide waves, we assume that each electron is represented by an ensemble of particles (walkers) and attached guide waves [15,16]. However, in dBB theory we have two related but not equivalent equations of motion for the particles, Eq. (12) and Eq. (13). In order to avoid the calculation of the quantum potential in Eq. (13) we choose to preserve the dBB theory as a first order theory, where the second-order Newtonian concepts of acceleration and force are disregarded. Therefore we keep the first order Eq. (12) as a guiding equation for the particles while Eq. (13) is used to reduce the 3$N$-dimensional Schrödinger equation to a set of coupled three-dimensional Schrödinger equations for the separate guide waves. To this end, we first represent the many-body classical potential in Eq. (1) as a sum of electron-nuclear and electron-electron parts:

$$V(\mathbf{r}_1,...,\mathbf{r}_N) = V_{e-n}(\mathbf{r}_1,...,\mathbf{r}_N) + V_{e-e}(\mathbf{r}_1,...,\mathbf{r}_N) = \sum_{k=1}^{N} V_{e-n}(\mathbf{r}_k) + \sum_{\substack{k,l=1 \\ k>l}}^{N} V_{e-e}(\mathbf{r}_k - \mathbf{r}_l) \qquad (18)$$

Then, the effective force calculated from the classical potential in Eq. (13) can be written as:

$$F_i^{cl}(\mathbf{r}_i) = -\nabla_i \left[ V_{e-n}(\mathbf{r}_i) + \sum_{\substack{j=1 \\ j \neq i}}^{N} V_{e-e}(\mathbf{r}_i - \mathbf{r}_j) \right] \qquad (19)$$

In order to make the many-body quantum equations more tractable we next reduce the quantum potential $Q(\mathbf{R},t)$ in Eq. (10) to a sum of quantum potentials for the separate particles by factorizing the amplitude of the many-body wave-function in Eq. (7) $\Psi(\mathbf{r}_1,...,\mathbf{r}_N,t) \approx R_1(\mathbf{r}_1,t)...R_N(\mathbf{r}_N,t)\exp[iS(\mathbf{r}_1,...,\mathbf{r}_N,t)/\hbar]$. Such representation can be motivated by the fact that in the de Broglie-Bohm theory the particle motion is determined by the gradient of the many-body phase function (see Eq. (12)) which is kept intact by the above representation. Thus we have:

$$R(\mathbf{r}_1,...,\mathbf{r}_N,t) = R_1(\mathbf{r}_1,t)...R_N(\mathbf{r}_N,t), \tag{20}$$

which, from Eq. (10), gives:

$$Q(\mathbf{R},t) = \sum_{i=1}^{N} Q_i(\mathbf{r}_i,t), \tag{21}$$

where:

$$Q_i(\mathbf{r}_i,t) = -\frac{\hbar^2}{2m}\frac{\nabla_i^2 R_i(\mathbf{r}_i,t)}{R_i(\mathbf{r}_i,t)} \tag{22}$$

is the quantum potential experienced by the $i$-th particle. Then, from Eq. (13) and from Eqs. (19)-(22), we obtain the 3D equation of motion for each individual particle:

$$m\frac{d^2\mathbf{r}_i}{dt^2} = \left\{-\nabla_i\left[-\frac{\hbar^2}{2m}\frac{\nabla_i^2 R_i(\mathbf{r}_i,t)}{R_i(\mathbf{r}_i,t)} + V_{e-n}(\mathbf{r}_i) + \sum_{\substack{j=1\\j\neq i}}^N V_{e-e}(\mathbf{r}_i - \mathbf{r}_j)\right]\right\}_{\substack{\mathbf{r}_i=\mathbf{r}_i(t)\\\mathbf{r}_j=\mathbf{r}_j(t)}}, \quad i,j=1,...,N \quad (23)$$

Next, it is assumed in TDQMC that each electron is described by a statistical ensemble of walkers and a separate ensemble of guide waves where each guide wave is attached to the corresponding walker. Then the guide wave $\varphi_i^k(\mathbf{r}_i,t)$ that is attached to the $k$-th walker from the $i$-th electron ensemble obeys the 3D time-dependent Schrödinger equation[15,16]:

$$i\hbar\frac{\partial}{\partial t}\varphi_i^k(\mathbf{r}_i,t) = \left[-\frac{\hbar^2}{2m}\nabla_i^2 + V_{e-n}(\mathbf{r}_i) + \sum_{\substack{j=1\\j\neq i}}^N V_{e-e}[\mathbf{r}_i - \mathbf{r}_j^k(t)]\right]\varphi_i^k(\mathbf{r}_i,t), \quad (24)$$

It can be easily verified that equation (23) is related to equation (24) through a standard polar decomposition, as in Eq. (7). It is important, however, that by solving Eq. (24) instead of Eq. (23) we avoid the explicit calculation of the quantum potential. The calculation of the quantum potential can pose a significant numerical problem for it is inversely proportional to the amplitude of the wave function and thus it becomes singular whenever the amplitude becomes small, e.g. near the nodes, see Eq. (22). It is noteworthy also that the continuity equation, Eq. (9), remains intact after the factorization done in Eq. (20). What we have ignored using that factorization is the contribution of the nonlocal quantum correlation effects which the many-body wave-function introduces on the particle motion. One such effect occurs due to the exchange interaction between the parallel spin electrons. While the third term in Eq. (24) accounts for the dynamic correlation between the electrons, the exchange interaction is accounted for in TDQMC through the guiding equation for the particles, Eq. (15), by representing the many-body quantum state as an antisymmetrized product (Slater determinant):

$$\Psi(\mathbf{r}_1,\mathbf{r}_2,...,\mathbf{r}_N,t) = A\prod_{i=1}^{N}\varphi_i(\mathbf{r}_i,t), \qquad (25)$$

where $A$ is the antisymmetrization operator which also includes the spin states of the particles. The use of Slater determinant for the $N$-dimensional quantum state introduces time-dependent nodal surfaces and pockets in the probability distribution that additionally rule the motion of the walkers through the drift velocity term (from Eq. (16)):

$$\mathbf{v}_{Di}^{k}(t) = \frac{\hbar}{m}\text{Im}\left[\frac{1}{\Psi(\mathbf{r}_1,...,\mathbf{r}_N,t)}\nabla_i\Psi(\mathbf{r}_1,...,\mathbf{r}_N,t)\right]_{\mathbf{r}_j=\mathbf{r}_j^k(t)} \qquad (26)$$

Whenever a walker approaches a nodal surface, the drift velocity in Eq. (26) grows and carries it away. The latter follows also from the general properties of quantum trajectories where the trajectories are not allowed to cross through nodal regions of the wave function where the phase becomes discontinuous and the probability of finding a particle should be zero. With other words, since the drift velocity in Eq. (4) and Eq. (15) is very large around nodal surfaces, the random walk is swept away as it approaches a node. Equations (14), (15), (24) - (26) comprise the complete set of equations of the TDQMC method. It is important to point out that the TDQMC method used here describes the system in terms of particle density without explicit reference to the many-body wave function. The calculation of $\Psi(\mathbf{R},t)$ in Eq. (25) gives result $\Psi^k(\mathbf{R},t)$ for each different set of guide waves $\varphi_i^k(\mathbf{r}_i,t)$ and is thus considered to be a statistical representative of the many-body quantum state. It is the walker density in configuration space obtained from TDQMC which reproduces the modulus square of the many-body wave function.

## 5. Time-dependent Hartree (-Fock) approximation within TDQMC

Time dependent Hartree (TDH) approximation can be obtained by following the familiar procedure of factorization of the many-body wavefunction as a product of single-particle functions $\Psi(\mathbf{r}_1,...,\mathbf{r}_N,t) = \varphi_1(\mathbf{r}_1,t)...\varphi_N(\mathbf{r}_N,t)$, and then substitute in Eq. (1):

$$i\hbar \frac{\partial}{\partial t}\varphi_i(\mathbf{r}_i,t) = \left[ -\frac{\hbar^2}{2m}\nabla_i^2 + V_{e-n}(\mathbf{r}_i) + \sum_{\substack{j=1 \\ j \neq i}}^{N} \int d\mathbf{r}_j V_{e-e}(\mathbf{r}_i - \mathbf{r}_j)\left|\varphi_j(\mathbf{r}_j,t)\right|^2 \right]\varphi_i(\mathbf{r}_i,t) \qquad (27)$$

Note that here we have used a complete factorization of the many-body quantum state, unlike in Eq. (20) where only its amplitude was factorized. If we now substitute the probability density for the *i*-th electron with its representation as a sum of delta-functions over an ensemble of Monte-Carlo sample points:

$$\left|\varphi_j(\mathbf{r}_j,t)\right|^2 = \frac{1}{M}\sum_{k=1}^{M}\delta\left[\mathbf{r}_j - \mathbf{r}_j^k(t)\right] \qquad (28)$$

we obtain:

$$i\hbar\frac{\partial}{\partial t}\varphi_i(\mathbf{r}_i,t) = \left[-\frac{\hbar^2}{2m}\nabla_i^2 + V_{e-n}(\mathbf{r}_i) + \sum_{\substack{j=1\\j\neq i}}^{N}\overline{V}_{e-e}[\mathbf{r}_i - \mathbf{r}_j(t)]\right]\varphi_i(\mathbf{r}_i,t), \qquad (29)$$

where:

$$\overline{V}_{e-e}[\mathbf{r}_i - \mathbf{r}_j(t)] = \frac{1}{M}\sum_{k=1}^{M}V_{e-e}[\mathbf{r}_i - \mathbf{r}_j^k(t)] \qquad (30)$$

is the average electron-electron potential seen by the *i*-th electron due to all Bohmian particles (walkers) which represent the *j*-th electron. In fact, Eq. (29) and Eq. (30) give the TDQMC version of the time-dependent Hartree approximation, where all walkers with coordinates $\mathbf{r}_i^k(t)$ are guided by the same function $\varphi_i(\mathbf{r}_i,t)$ through the equation:

$$\frac{d\mathbf{r}_i^k}{dt} = \frac{\hbar}{m}[\alpha\,\text{Re} + \text{Im}]\left[\frac{1}{\varphi_i(\mathbf{r}_i,t)}\nabla_i\varphi_i(\mathbf{r}_i,t)\right]_{\mathbf{r}_i=\mathbf{r}_i^k(t)}, \qquad (31)$$

It is seen from Eqs. (29)-(31) that in the TDH approximation the walker motion in not strictly correlated since all guiding waves in Eq. (29) depend on an average electron-electron potential. Since in this case the probability distribution of the ensemble of trajectories $\mathbf{r}_i^k(t)$ reproduces the modulus square of the wave-function $|\varphi_i(\mathbf{r}_i,t)|^2$, the solution of the coupled Eqs. (29)-(31) gives the same result as the direct solution of Eq. (27). Nevertheless, for many-electron problems the use of Monte Carlo approach to time-dependent Hartree approximation can be advantageous because it replaces the integral in Eq. (27) by a Monte Carlo sum in Eq. (30), which can be calculated very efficiently.

The time dependent Hartree-Fock (TDHF) approximation can be considered in a similar manner. Substituting Eq. (25) into Eq. (1) yields:

$$i\hbar \frac{\partial}{\partial t}\varphi_i(\mathbf{r}_i,t) = \left[-\frac{\hbar^2}{2m}\nabla_i^2 + V_{e-n}(\mathbf{r}_i) + \sum_{\substack{j=1 \\ j\neq i}}^{N}\int d\mathbf{r}_j V_{e-e}(\mathbf{r}_i-\mathbf{r}_j)|\varphi_j(\mathbf{r}_j,t)|^2\right]\varphi_i(\mathbf{r}_i,t)$$

$$-\sum_{\substack{j=1 \\ j\neq i}}^{N}\left\{\int d\mathbf{r}_j V_{e-e}(\mathbf{r}_i-\mathbf{r}_j)\varphi_i(\mathbf{r}_j,t)\varphi_j^*(\mathbf{r}_j,t)\right\}\varphi_j(\mathbf{r}_i,t) \qquad (32)$$

The direct (Coulomb) term in Eq. (32) can be calculated using Monte-Carlo integration over the walker distribution. However, this cannot be done easily for the exchange term, and numerical integration in coordinate space must be performed instead. One important advantage of the fully correlated TDQMC method is that the calculation of exchange integrals is avoided since the Coulomb interaction is separated from the exchange interaction and the latter is accounted for via the guiding equation, Eq. (15). In TDHF (Eq. (32)) the evolution of the wavefunctions is determined by potentials that are averaged over all walkers in both the direct and the exchange terms. As a result, the correlation effects are washed out.

## 6. Electron density and energy estimation

In order to find the total energy of a system of $N$ electrons, we take the average over ensemble of $M$ Bohmian particles (walkers) which represent each electron. If we assume that there is no random motion in steady-state, from Eq. (12) and Eq. (8) we obtain:

$$E = \frac{1}{M} \sum_{k=1}^{M} \left[ \sum_{i=1}^{N} \left[ \frac{1}{2} m \dot{\mathbf{r}}_i^{k\,2} + Q(\mathbf{r}_1,...,\mathbf{r}_i^k,...,\mathbf{r}_N,t) + V_{e-n}(\mathbf{r}_i^k,t) \right] + \sum_{\substack{i,j=1 \\ i>j}}^{N} V_{e-e}(\mathbf{r}_i^k - \mathbf{r}_j^k) \right], \qquad (33)$$

where the irreducible quantum potential has been estimated for the trajectory $\mathbf{r}_i = \mathbf{r}_i^k(t)$:

$$Q_i(\mathbf{r}_1,...,\mathbf{r}_i^k,...,\mathbf{r}_N,t) = \left[ -\frac{\hbar^2}{2m} \frac{\nabla_i^2 R(\mathbf{r}_1,...,\mathbf{r}_i,...,\mathbf{r}_N,t)}{R(\mathbf{r}_1,...,\mathbf{r}_i,...,\mathbf{r}_N,t)} \right]_{\mathbf{r}_i = \mathbf{r}_i^k(t)} \qquad (34)$$

A simple estimation for the energy of the system in stationary state at instant $\tau$ can be obtained by setting the velocities of all walkers to be zero ($\dot{\mathbf{r}}_i^k = 0$), which from Eq. (16) yields $S(\mathbf{R},\tau) = const$. Then, using the factorization $R(\mathbf{r}_1,...,\mathbf{r}_N,\tau) = \varphi_1(\mathbf{r}_1,\tau)...\varphi_N(\mathbf{r}_N,\tau)$, from Eqs. (10), (22) and (33), we get: [16]

$$E = \frac{1}{M} \sum_{k=1}^{M} \left[ \sum_{i=1}^{N} \left[ -\frac{\hbar^2}{2m} \frac{\nabla_i^2 \varphi_i^k(\mathbf{r}_i^k)}{\varphi_i^k(\mathbf{r}_i^k)} + V_{e-n}(\mathbf{r}_i^k) \right] + \sum_{\substack{i,j=1 \\ i>j}}^{N} V_{e-e}(\mathbf{r}_i^k - \mathbf{r}_j^k) \right]_{\substack{\mathbf{r}_i^k = \mathbf{r}_i^k(\tau) \\ \mathbf{r}_j^k = \mathbf{r}_j^k(\tau)}} \qquad (35)$$

Although Eq. (35) gives energies that are very close to the exact values [16], a small portion of the non-local correlation energy still remains neglected because the irreducible quantum potential depends, in general, on the shape of $R(\mathbf{r}_1,...,\mathbf{r}_N,\tau)$ in configuration space. Therefore, the energy estimate can be improved further if we first evaluate the function $R(\mathbf{r}_1,...,\mathbf{r}_N,\tau)$, and then make use of Eq. (10) to calculate the quantum potential. To this end, we first obtain a smoothed (differentiable) probability density function which represents the density of the walkers for each electron. Various methods have been employed to estimate the density of points in quantum hydrodynamics. Maddox and Bittner [22] have used iterative procedure to find the parameters of a set of Gaussians that best approximate the density function. Grashchuk and Rassolov [45] used a global method, least squares fitting, to approximate the electron density as a sum of Gaussians. Here we employ a nonparametric method, kernel density estimation (KDE), in order to estimate the quantum potential, and hence to obtain the steady state energy of the quantum system.

First, we note that the problem of finding the probability density distribution of a set of points in multidimensional space represents one of the basic problems in data mining. In KDE method [46], also known as smoothed particle hydrodynamics method [47], the density at point $\mathbf{R}$ of an ensemble of $M$ Monte-Carlo points is usually estimated using Gaussian kernels:

$$f(\mathbf{R}) = \frac{1}{M|\Sigma|^{0.5}(2\pi h)^{D/2}} \sum_{i=1}^{M} \exp\left[-\frac{(\mathbf{R}-\mathbf{R}_i)^T \Sigma^{-1}(\mathbf{R}-\mathbf{R}_i)}{2h^2}\right], \tag{36}$$

where $D$ is the dimensionality of the configuration space, $\Sigma$ is the $D$x$D$ "orientation" matrix which is equal to the covariance matrix of the data, divided by $h^2$. Here $h$ is the scaling factor (bandwidth) which determines the width of the Gaussians in Eq. (36). For simple static kernels, the parameters of the Gaussians are the same for all points and the bandwidth is given by Silverman's rule-of-thumb [46]:

$$h \propto \sigma M^{-1/(D+4)}, \tag{37}$$

where σ is the standard deviation for the whole Monte Carlo sample. Since the covariance matrix reflects the symmetry of the set of Monte Carlo sample points with respect to the axes in configuration space, that matrix can be diagonalized by rotation to principal axes. Then, the different bandwidths are determined from the diagonal elements of the resulting matrix that usually improves the density estimation. An alternative is to use product kernel estimator of the following form [47]:

$$f(\mathbf{R}) = \frac{1}{M h_1 \ldots h_N} \sum_{i=1}^{M} \left\{ \prod_{j=1}^{D} \exp\left[ -\frac{(\mathbf{R} - \mathbf{R}_i)^2}{2 h_j^2} \right] \right\} \tag{38}$$

where the bandwidth $h_j$ depends on the standard deviation of the whole sample along the $j$-th axis in configuration space [48]:

$$h_j = \left( \frac{4}{(D+2)M} \right)^{1/(D+4)} \sigma_j \tag{39}$$

However, since $\sigma_j$ is a global quantity for the whole particle ensemble, large errors for significant variations of the density can be expected. Therefore, adaptive KDE is used where the bandwidth becomes a local quantity which reflects the fact that in regions of high density one can estimate the local distribution with narrower Gaussians, and vice versa. It can be shown through nearest neighbor approach that asymptotically we have $h_j \propto 1/\sqrt{f(R_j)}$ [49]. Thus, the adaptive KDE can be considered to be a second iteration with respect to the fixed bandwidth estimation technique.

A major advantage of using KDE for estimation of the quantum potential is that the second derivatives in Eq. (10) can be calculated without referencing to finite differences of multi-variate functions. Instead, the derivatives are calculated via analytical differentiation of the kernel function (Gaussians in Eqs. (36) and (38)). Once the particle probability density function $P(\mathbf{R}, \tau)$ and its second derivative is found, the total energy of the many-electron system in stationary state can be estimated from the formula:

$$E = \int P(\mathbf{R},\tau) \left[ \frac{\hbar^2}{8m} \frac{(\nabla P)^2}{P^2} + V(\mathbf{R}) \right] d\mathbf{R} \tag{40}$$

The integration in Eq. (40) is easily performed using Monte-Carlo sum over the particle distribution:

$$E = \frac{1}{M} \sum_{k=1}^{M} \left[ \frac{\hbar^2}{8m} \sum_{i=1}^{N} \frac{(\nabla_i P)^2}{P^2} + \sum_{i=1}^{N} V_{e-n}(\mathbf{r}_i^k) + \sum_{\substack{i,j=1 \\ i>j}}^{N} V_{e-e}(\mathbf{r}_i^k - \mathbf{r}_j^k) \right]_{\substack{\mathbf{r}_i^k = \mathbf{r}_i^k(\tau) \\ \mathbf{r}_j^k = \mathbf{r}_j^k(\tau)}} \tag{41}$$

## 7. Algorithm

The TDQMC algorithm involves numerical solution of a set of coupled equations for the guide waves, Eq. (24), and the trajectory equations, Eqs. (14), (15), together with the symmetry condition, Eq. (25). Let us recall that in TDQMC the walker distribution corresponds to $|\Psi(\mathbf{R})|^2$ and therefore the problem with the negative values of $\Psi(\mathbf{R})$ (the fermion sign problem in DMC) is not present.

First, the ground state of the quantum system is calculated by propagating the initial set of guide waves $\varphi_j^k(\mathbf{r}_j, t=0)$, usually Gaussians, in complex time until steady state in electron energy (Eq. (35) or Eq. (41)) is established. The use of complex time $t = t' + it''$ in Eq. (24) ensures that the guide waves relax to the ground state owning to $t''$ while each of these waves acquires a time-dependent phase due to $t'$, which guides the walkers to stationary positions through Eq. (14). Since at steady state the velocity of the walkers tends to zero, the amplitude α of the random component in Eqs. (14) and (15) is assumed to be a decreasing function of time [16]. In fact, the random component in Eq. (14) causes thermalization of the particle ensemble at each time step that is needed to avoid possible bias in the walker distribution that may arise due to the quantum drift alone. An alternative way to achieve thermalization is to set $\alpha=0$ in Eqs. (14), (15) and use Metropolis algorithm instead to sample the densities $|\varphi_i^k(\mathbf{r}_i, t)|^2$ at each time step [16]. Once steady state is established, the imaginary time component $t''$ is set to zero, and the evolution of the system proceeds in real time for both guide waves and particles where the random component in particle motion may be reduced significantly. It should be noted that the system evolution towards steady state is not ergodic, and therefore all ensemble averages have to be calculated for the final particle distribution.

The TDQMC algorithm can be summarized in the following steps:

1. Generate an initial set of guide waves $\varphi_i^k(\mathbf{r}, t=0)$ for each separate walker, where $i=1,...,N$ denotes the electron, and $k=1,...,M$ denotes the walkers associated with that electron. All initial guide waves can be Gaussians of width $\sigma$, centered at the origin of the coordinate system.

2. Generate an initial ensemble of walkers at random positions $\mathbf{r}_i^k(t=0)$ in physical space, with Gaussian distribution of width $\sigma$. Alternatively, the initial ensemble can be generated using Metropolis algorithm which samples the probability distribution obtained from Hartree (- Fock) approximation.

3. Stochastic step: move each walker from position $\mathbf{r}_i^k$ to position $\mathbf{r}_i^{k'}$ according to

$$\mathbf{r}_i^{k'} = \mathbf{r}_i^k + \mathbf{\eta}\sqrt{\frac{\alpha\hbar}{m}dt'}, \tag{42}$$

or use Metropolis sampling for $\alpha=0$.

4. Calculate the guide waves at instant $t+dt$ from Eq. (24).

5. Drift step: move each walker from position $\mathbf{r}_i^{k'}$ to position $\mathbf{r}_i^{k''}$ according to:

$$\mathbf{r}_i^{k''} = \mathbf{r}_i^{k'} + \mathbf{v}(\mathbf{r}_i^{k'})dt', \tag{43}$$

where:

$$\mathbf{v} = \frac{\hbar}{m}[\alpha\,\mathrm{Re}+\mathrm{Im}]\left[\frac{1}{\Psi(\mathbf{r}_1,...,\mathbf{r}_N,t+dt)}\nabla_i\Psi(\mathbf{r}_1,...,\mathbf{r}_N,t+dt)\right]_{\mathbf{r}_j=\mathbf{r}_j^k(t)}, \tag{44}$$

and the anti-symmetry of the wavefunction has been taken explicitly into account through Eq. (25).

6. Calculate the energy of the stationary states from Eq. (35) or Eq. (41).

7. Switch to real time in Eq. (24), and turn on any external fields to study time dependent dynamics. Since the algorithm for calculation of the ground state of the system guaranties that $P(\mathbf{R},\tau) = |\Psi(\mathbf{R},\tau)|^2$, it follows from the continuity equation, Eq. (9), that $P(\mathbf{R},t) = |\Psi(\mathbf{R},t)|^2$ will hold at all times during the evolution.

In the above algorithm, split-time-step approach has been implemented for the motion of the walkers, where the random (diffusive) and the guiding (quantum drift) components are separated by a single time step for updating the guiding waves. It is assumed that for sufficiently small time step both Coulomb and exchange interactions between the electrons can be accounted for accurately. During the calculation of the ground state the guide waves are normalized at each time step.

## 8. Numerical results

To illustrate the performance of TDQMC method described in the previous sections we calculate the ground state wavefunction and the time-dependent dipole moment of one-dimensional Helium atom. This model atom has proven to be very useful in modeling the interaction of atomic systems with intense ultashort laser pulses (e.g. in [50]). The model employs smoothed Coulomb potentials to avoid numerical complications from the singularity at the origin, which also allows a fine adjustment of the ground state energy of the atom. Here we assume that the electron-nuclear and the electron-electron interactions are described by the following potentials:

$$V_{e-n}(x_i) = -\frac{2e^2}{\sqrt{a+x_i^2}} ; \qquad (45)$$

$$V_{e-e}[x_i - x_j^k(t)] = \frac{e^2}{\sqrt{b+[x_i - x_j^k(t)]^2}} , \qquad (46)$$

where $i=1,2$; $k=1,...,M$, and $a$ and $b$ are smoothing parameters. In order to demonstrate the role of the electron correlation on the ground state shape and energy, we have chosen $a=1$ a.u. (atomic units) and $b=0.2$ a.u. in Eqs (45), (46). The ground state of the model atom is calculated by initially choosing an ensemble of $M=5000$ Bohmian particles to serve as random walkers in our Monte Carlo simulation. For atom in a spin-singlet ground state, the two-body wavefunction in Eq. (1) is a symmetrized product of two one-electron orbitals, similarly to the unrestricted TDHF model. Next, we assign a separate guide function $\varphi_i^k(x_i,t)$ to each Bohmian particle (walker) with coordinate $x_i(t)$ to guide the particle motion in accordance with Eqs. (42)-(44). The initial distributions of the particles (walkers) that represent the two electrons are Gaussians with standard deviation $\sigma=1$ a.u. After propagation over 300 complex time steps in Eq. (24), the initial ensembles evolve towards steady state with ground state energy -1.936 a.u. obtained from Eq. (35), and -1.940 a.u. from Eq. (41). The exact ground state energy found from a direct diagonalization of the atomic Hamiltonian is -1.941 a.u. which is very close to the TDQMC result, while the Hartree-Fock approximation gives -1.834 for the ground state energy. The walker distribution and the corresponding probability density function obtained from Eq. (38) for a symmetric (spin-singlet) ground state are depicted in Figure 1(a),(b), respectively. On the other hand, for parallel spin electrons, the lowest energy state of the model atom is an anti-symmetric function under exchange of the electrons (fermionic ground state). In this case the guiding waves which belong to different electrons are orthogonalized using Gram-Schmidt procedure. The walkers' distribution and the smoothed probability density in this case are shown in Fig.1(c),(d).

The ground state energy estimate from TDQMC calculation is -1.766 a.u. which is very close to the exact result -1.768 a.u.

In order to calculate the response of the model atom to an external time varying fields we first perform a gauge transformation $\hat{\mathbf{p}} \to \hat{\mathbf{p}} - e/c\mathbf{A}$ in the time dependent Schrödinger equation (24) and in the guiding equation (44), where $\hat{\mathbf{p}}$ is the momentum operator and $A$ is the vector potential of the external electromagnetic field. Then, the time response of 1D Helium atom can be estimated by calculating its induced dipole moment as function of time. The ensemble average atomic dipole moment is calculated as a sum of the dipole moments of the different walkers through the formula:

$$d(t) \propto \sum_{i=1}^{2} \sum_{k=1}^{M} \int \varphi_i^{k*}(x,t) x \varphi_i^k(x,t) dx \qquad (47)$$

Our goal here is to compare the TDQMC result for the time-dependent dipole moment with the result from the direct solution of TDSE in two-dimensional configuration space ("exact" solution), and from TDHF approximation, for the same set of parameters. For atom in a singlet ground state, with $a$=1, $b$=1.5 a.u. in Eqs. (45), (46), we choose an electromagnetic pulse with duration 0.5 femtoseconds at wavelength $\lambda$=57 nm, and peak intensity 5.05 $10^{14}$ W/cm$^2$. Figure 2 depicts the time profile of that pulse. In Fig.3 we show the result for the dipole moment from TDQMC calculation (solid line), compared with the direct numerical integration of 2D time-dependent Schrödinger equation (dashed line), and from TDHF approximation (dotted line). It is seen that the prediction of TDQMC and the "exact" result are very close while they differ significantly from the TDHF result. That difference increases especially for later times where the electron correlation in one spatial dimension enhances the ionization of the atom and de-phases the electron oscillations, which results in inhibited oscillations of the dipole moment. The time dependent behavior of the dipole moment for 1D Helium atom in an antisymmetric ground state is similar. In order to get more detailed information about the dipole response of the atom when irradiated by a strong laser pulse we calculate the Fourier transform of the dipole acceleration and plot in Fig.4 the harmonic spectra for the TDQMC, the "exact" solution, and the TDHF approximation. Since the higher-order harmonics are very sensitive to the accuracy of the approximation used, it is seen from Fig. 4 that the TDQMC prediction is very close to the exact result while they differ significantly from the TDHF prediction. The results presented in Fig.3 and Fig.4 indicate that the electron correlation effects are accurately taken into account by the TDQMC technique for the case of atom interacting with external electromagnetic field. Further examples of TDQMC predictions can be found in [51].

## 9. Conclusions

In this paper we have compared the method of stochastic quantization and the newly developed time dependent quantum Monte Carlo approach where quantum dynamics is modeled using ensembles of particles and guiding waves. It is shown that both TDQMC and stochastic quantization use similar guiding equations for the Bohmian particles (walkers). However, the motion of both particles and guide waves occurs in physical space in TDQMC while it occurs in configuration space in stochastic quantum mechanics. Also, the use of Metropolis sampling in TDQMC is often more advantageous than using osmotic velocity (as in stochastic dBB method) for sampling the quantum probability density. In TDQMC the many-body time dependent Schrödinger equation is reduced to a set of coupled one-body Schrödinger equations for the separate guiding waves. The role of the particles (walkers) is to accomplish the connection between the different guide waves in a self-consistent manner. TDQMC offers significant advantages over the standard method of diffusion quantum Monte Carlo in that in TDQMC the guide waves evolve in time together with particles' trajectories and, therefore, no preliminary knowledge of the nodes of the many-body wavefunction is required. Also, in TDQMC the walker distribution corresponds to modulus square of the wavefunction and the problem with the interpretation for fermionic states is avoided. Hartree approximation can be naturally interpreted in terms of particle ensembles and guide waves for non-correlated electrons. On the other hand, the nonlocal correlation effects enter TDQMC calculation through the guiding equation where the particle velocity is expressed in terms of symmetric or anti-symmetric products of individual guide waves. Thus the separate calculation of the dynamic correlation (that is due to the Coulomb interaction between the electrons) and the exchange-induced correlation that we use in TDQMC eliminates the need of generally unknown exchange-correlation potentials.

**Acknowledgments**

The author gratefully acknowledges support from the National Science Fund of Bulgaria under contract WUF-02-05.

**Figure captions:**

**Figure 1**. Walker distribution in configuration space for 1D Helium in steady state: (a) for symmetric (singlet-spin) ground state, (c) for antisymmetric (triplet spin) ground state. The contour maps in (b) and (d) depict the smoothed probability density as a result of kernel density estimation of the distributions in (a) and (c), respectively.

**Figure 2**. Time dependence of the electric field used in the calculation of the dipole moment.

**Figure 3**. Time dependence of the dipole moment of 1D Helium from TDQMC method (solid line), exact solution (dashed line), and TDHF approximation (dotted line). Pulse duration 0.5 femtoseconds, wavelength $\lambda$=57 nm, peak intensity 5.05 $10^{14}$ W/cm$^2$. Singlet atomic ground state with *a*=1 a.u. and *b*=1.5 a.u. is assumed.

**Figure 4.** Harmonic spectra (logarithmic scale) obtained from 1D Helium atom for pulse duration 1 femtosecond, wavelength $\lambda$=230 nm, peak intensity 3.16 $10^{15}$ W/cm$^2$. Singlet atomic ground state with *a*=1 a.u. and *b*=1.5 a.u. is assumed. Solid line - TDQMC result, dashed line – exact solution, dotted line – TDHF approximation.

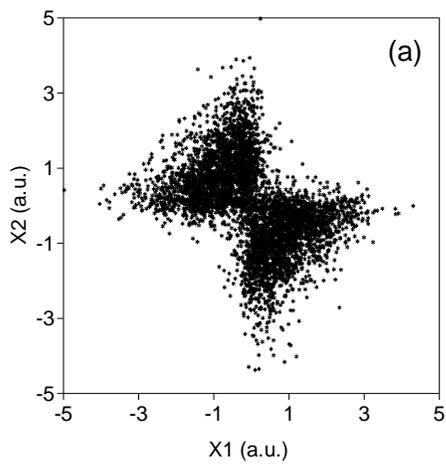
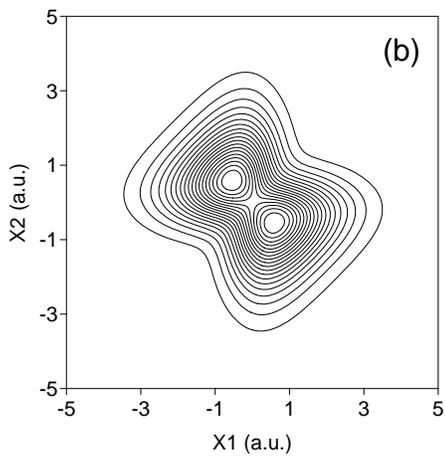
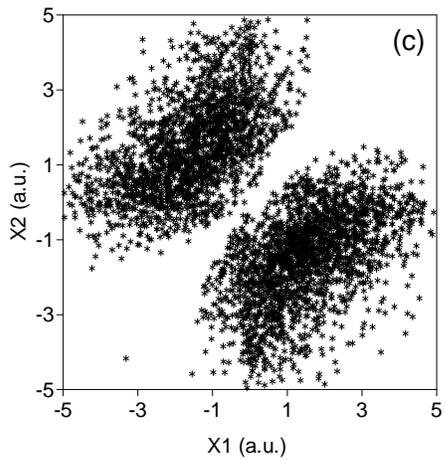
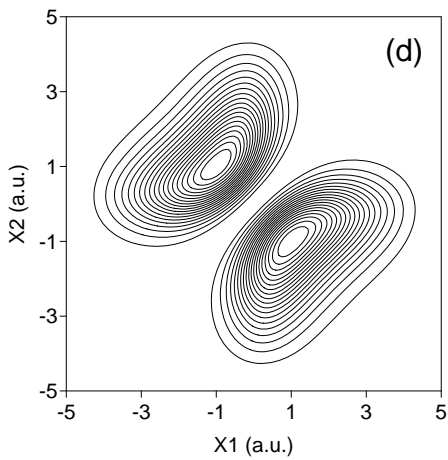

I. Christov, Fig.1

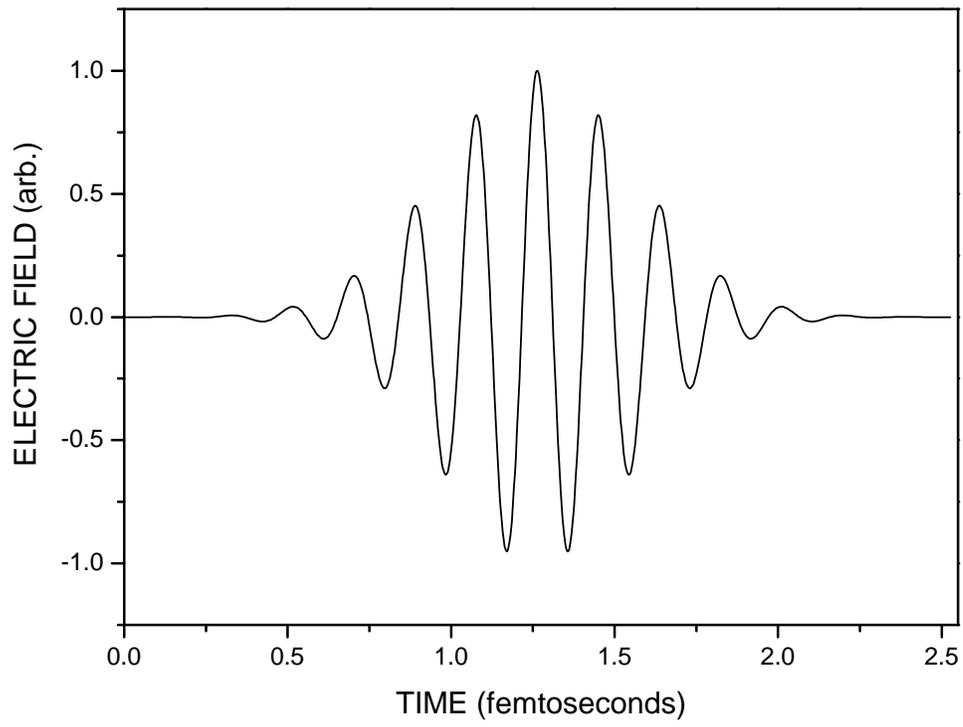

I. Christov, Fig.2

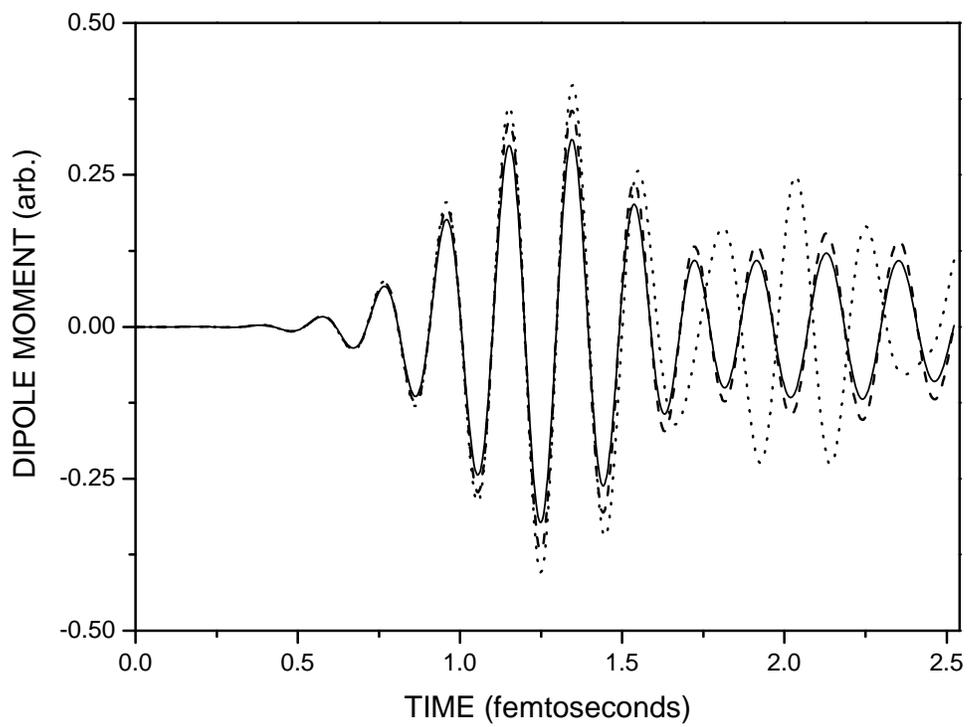

I. Christov, Fig.3

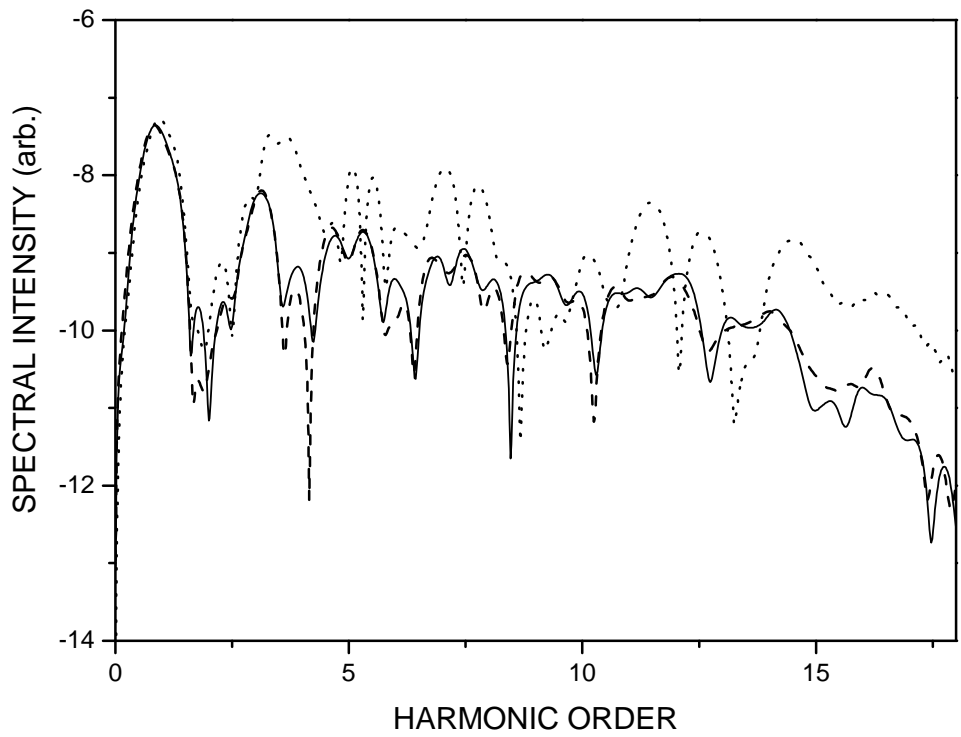

I. Christov, Fig.4